\documentclass[
twocolumn,
amsmath, amssymb, amsfonts,preprintnumbers,aps,prd,longbibliography,nofootinbib]{revtex4-2}

\usepackage{graphicx}
\pdfoutput=1

\usepackage{dcolumn}
\usepackage{bm}
 \usepackage{amstext}
 \usepackage{amssymb}
 \usepackage{amsmath}

 \usepackage{color}
 \usepackage{bbold}
 \usepackage{delimset} 
\usepackage[caption=false,justification=justified]{subfig}
\usepackage[colorlinks=true]{hyperref}
\usepackage{orcidlink}
\usepackage{float}
\usepackage[normalem]{ulem}
\usepackage{mathtools}
\usepackage[colorlinks=true]{hyperref}

\usepackage{tikz}
\usetikzlibrary{decorations.pathmorphing}
\usetikzlibrary{decorations.markings}
\usetikzlibrary{positioning, shapes, snakes}
\usetikzlibrary{arrows.meta}

\begin{document}

\author{Yong Zhang \!\orcidlink{0000-0002-3522-0885}}

\email{zhangyong1@nbu.edu.cn}

\affiliation{
Institute of Fundamental Physics and Quantum Technology\\ \& School of Physical Science and Technology, Ningbo University, Ningbo, Zhejiang 315211, China }

\title{A complex-linear reformulation of Hamilton--Jacobi theory
\\ 
and emergent quantum structure}

\begin{abstract}

Classical mechanics admits multiple equivalent formulations, from Newton's equations to the variational Lagrange--Hamilton framework and the scalar Hamilton--Jacobi (HJ) theory.  In the HJ formulation, classical ensembles evolve through the continuity equation for a real density $\rho = R^{2}$ coupled to Hamilton's principal function $S$. Here we develop a complementary formulation---the Hamilton--Jacobi--Schr\"odinger (HJS) theory---by embedding the pair $(R,S)$ into a single complex field.  Starting from a completely general complex ansatz $\psi = f(R,S)\, e^{i g(R,S)},$ and imposing two minimal structural requirements, we obtain a unique map $\psi = R\, e^{iS/\kappa}$ together with a linear HJS equation whose $|\kappa| \to 0$ limit reproduces the HJ formulation exactly.  Remarkably, when $\mathrm{Re}(\kappa)\neq 0$, essential features of quantum mechanics---superposition, operator algebra, commutators, the Heisenberg uncertainty principle, Born's rule and unitary evolution---follow naturally as structural consistency conditions.  HJS thus provides a unified mathematical viewpoint in which classical and quantum dynamics appear as different limits of a single underlying structure.

\end{abstract}
 
\maketitle

\section*{Introduction}

Classical mechanics admits several structurally distinct formulations.
Newton's equations describe motion through forces, while the Lagrange--Hamilton
variational framework identifies physical trajectories by stationary action.
HJ theory provides yet another viewpoint: it recasts dynamics
into a pair $(\rho,S)$ consisting of a classical ensemble density $\rho$ obeying a
continuity equation and a Hamilton principal function $S$ satisfying a nonlinear
first-order partial differential equation
(see, e.g., \cite{LandauMechanics}). 
Although equivalent, these formulations emphasize different aspects of dynamics and
have shaped developments ranging from analytical mechanics to field theory and
general relativity.

In this work we revisit the HJ  formulation and ask whether its 
coupled evolution equations admit a linear representation.  
To this end, we consider the most general complex embedding of the real variables
$(R,S)$,
\[
  \psi = f(R,S)\, e^{i g(R,S)},
\]
with $\rho = R^{2}$ and $S$ the action function.  
Without invoking any quantum structure, we impose two minimal requirements 
motivated solely by linearization: (i) the evolution equation for $\psi$ must 
remain first order in time, and (ii) nonlinear terms such as 
$(\nabla\psi/\psi)^{2}$ must be absent.  
These mild conditions dramatically restrict the admissible complexifications and 
select a single minimal form,
\[
  \psi = R\, e^{iS/\kappa},
\]
where $\kappa$ is an arbitrary complex parameter.

Substituting the minimal complex embedding into the HJ  system
modifies the HJ equation itself.  
One obtains a deformed equation,
\[
\partial_t S + \frac{(\nabla S)^2}{2m} + V + Q[R] = 0,
\qquad
Q[R] = -\,\frac{\kappa^{2}}{2m}\,\frac{\nabla^{2}R}{R},
\]
The accompanying continuity equation,
\[
\partial_t (R^{2})
+ \nabla\!\cdot\!\left(R^{2}\,\frac{\nabla S}{m}\right)=0 ,
\]
is unchanged.  
This conservation law is an intrinsic part of the HJ framework and
remains exactly preserved for all values of $\kappa$.

Taken together, the deformed HJ equation and the classical 
continuity equation constitute the HJS system.  
On the support of the map $\psi = R e^{iS/\kappa}$, the pair recombines 
exactly into the linear equation
\[
i\kappa\,\partial_t\psi 
= -\frac{\kappa^{2}}{2m}\,\nabla^{2}\psi + V\psi ,
\]
so that a Schr\"odinger-type structure appears whenever $\mathrm{Re}(\kappa)\neq 0$, taking a linear form familiar from standard quantum mechanics 
(see, e.g., \cite{LandauQM}). 
In the classical limit $|\kappa|\to 0$, the deformation term $Q[R]$ disappears wherever the polar variables remain regular, and the HJS system reduces to standard HJ theory.

A key consequence is that the nonlinear HJ description is recovered
as a limiting case of a \emph{linear} evolution equation for~$\psi$.
This linearization provides a new functional representation of classical dynamics,
allowing tools normally reserved for wave equations---linear state-space methods, spectral
methods, and amplitude-based techniques---to be applied directly to classical
problems. Many classical systems acquire a more transparent global organization once expressed in this linear form.
Even in the familiar $1/r$ two-body problem,
the geometric orbit equation is simple, but the HJ dynamics and the
time evolution governed by Kepler's equation are implicit and not readily invertible;
by contrast, the corresponding Schr\"odinger description is algebraically linear
and admits closed-form solutions. The HJS map clarifies this contrast: the
apparent simplicity of quantum formulations reflects the underlying linearization
of HJ flow.

Such linear structures are already implicit in several modern developments \cite{Goldberger:2004jt,Damour2016PMEOB,Bern:2019nnu,Kalin:2020fhe}.
High-precision post-Minkowskian analyses of gravitational dynamics, for example,
rely heavily on Fourier and amplitude-based methods whose effectiveness stems
from representing classical motion through wave-like structures.
The HJS framework makes this structural feature explicit: it provides a
genuinely classical linearization in which the tools of spectral analysis,
superposition, and scattering amplitudes acquire a natural and unified place.

A useful analogy is provided by the ancient method of exhaustion for computing
the area of a circle.
The exact circular area is not obtained by treating the circle itself in its
full nonlinear form, but by replacing it with a family of simpler linear pieces
whose widths are taken to zero at the end.
In a similar spirit, the HJS formulation does not replace the classical HJ
dynamics by something unrelated; rather, it reorganizes a nonlinear problem
into a linear Schr\"odinger-type representation that is often far easier to
analyze.
The parameter $\kappa$ plays a role analogous to the strip width in such a
construction: as $|\kappa|\to 0$, the linear representation recovers the exact
classical HJ system, as schematically illustrated in Fig.~\ref{fig:exhaustion}.
From this viewpoint, the Schr\"odinger-type equation may be understood not only
as a hallmark of quantum theory, but also as an efficient intermediate
representation for classical dynamics.

\begin{figure}[t]
\centering
\begin{tikzpicture}[>=stealth, line cap=round, line join=round]

\def\xL{0.0}      
\def\xR{5.05}     
\def\yTop{0.95}   
\def\yBot{-2.95}  

\def\R{1.35}
\def\nstrips{19}
\pgfmathsetmacro{\dx}{2*\R/\nstrips}

\node at (\xR,2.55) {\small strip width $\to 0$};

\draw[thick] (\xL,\yTop) circle (\R);

\foreach \k in {1,...,18} {
    \pgfmathsetmacro{\x}{-\R + \k*\dx}
    \pgfmathsetmacro{\yy}{sqrt(max(0,\R*\R-\x*\x))}
    \draw[thick] (\xL+\x,\yTop-\yy) -- (\xL+\x,\yTop+\yy);
}

\draw[->, thick] (\xL+1.65,\yTop) -- (\xR-1.65,\yTop);

\def\nrect{19}
\def\w{0.12}
\def\gap{0.025}

\pgfmathsetmacro{\Span}{\nrect*\w + (\nrect-1)*\gap}
\pgfmathsetmacro{\RR}{\Span/2}
\pgfmathsetmacro{\xstart}{\xR-\Span/2}

\foreach \k in {1,...,19} {
    \pgfmathsetmacro{\xl}{\xstart + (\k-1)*(\w+\gap)}
    \pgfmathsetmacro{\xm}{\xl + 0.5*\w}
    \pgfmathsetmacro{\dist}{\xm-\xR}
    \pgfmathsetmacro{\h}{2*sqrt(max(0,\RR*\RR-\dist*\dist))}
    \draw[thick] (\xl,\yTop-0.5*\h) rectangle ++(\w,\h);
}

\pgfmathsetmacro{\xM}{(\xL+\xR)/2}
\node at (\xM,-0.72) {\Large $\Big\Downarrow$};
\node[anchor=west] at (\xM+0.40,-0.77) {\small in the same spirit};


\node at (\xR,-1.63) {\small $|\kappa| \to 0$};

\def\boxLW{3.15}
\def\boxLH{2.10}
\draw[thick, rounded corners=7pt]
(\xL-0.5*\boxLW,\yBot-0.5*\boxLH) rectangle
(\xL+0.5*\boxLW,\yBot+0.5*\boxLH);
\node[align=center] at (\xL,\yBot+0.10) {%
\small nonlinear HJ\\[-1pt]
\small equation\\[3pt]
{\footnotesize$\displaystyle \partial_t S + \frac{(\nabla S)^2}{2m} + V = 0$}
};

\draw[->, thick] (\xL+1.85,\yBot) -- (\xR-2.05,\yBot);

\def\boxRW{3.85}
\def\boxRH{2.10}
\draw[thick, rounded corners=7pt]
(\xR-0.5*\boxRW,\yBot-0.5*\boxRH) rectangle
(\xR+0.5*\boxRW,\yBot+0.5*\boxRH);
\node[align=center] at (\xR,\yBot+0.10) {%
\small linear\\[-1pt]
\small Schr\"odinger-type\\[-1pt]
\small equation\\[3pt]
{\footnotesize $\displaystyle i\kappa\,\partial_t\psi
=
-\frac{\kappa^{2}}{2m}\,\nabla^{2}\psi
+V\psi$}
};

\end{tikzpicture}
\caption{
Schematic analogy with the method of exhaustion.
A circle may be approximated by thin strips, each treated as a rectangle,
with the exact result recovered as the strip width tends to zero.
In the same spirit, the HJS formulation reorganizes nonlinear HJ dynamics
into a linear Schr\"odinger-type representation, with the exact classical
HJ system recovered as $|\kappa|\to 0$.
}
\label{fig:exhaustion}
\end{figure}

When $\mathrm{Re}(\kappa)\neq 0$, the linear HJS equation supports a richer
algebraic structure than its classical counterpart.
This extended structure brings in elements absent from HJ theory but central to quantum mechanics:
a linear state space, operator generators, and nontrivial commutators.
Once probabilistic consistency is imposed---leading to the Born rule and unitary
time evolution---the standard Heisenberg uncertainty principle follows in its
usual form.
Thus the core algebraic and probabilistic framework of quantum mechanics
appears as a natural extension of the HJS structure rather than an
independent set of postulates.

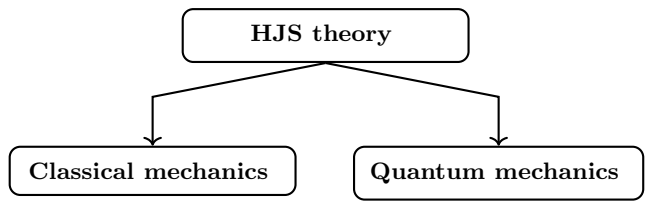
\begin{figure}[t]
\centering

\begin{tikzpicture}[
    box/.style={
        rectangle, draw, rounded corners,
        align=center,
        minimum width=3.8cm,
        inner sep=6pt,
        thick
    },
    arrow/.style={->, thick}
]

\node[box] (HJ) {
    \textbf{HJS theory}
};

\node[box, below=1.1cm of HJ, xshift=-2.3cm] (CL) {
    \textbf{Classical mechanics}
};

\node[box, below=1.1cm of HJ, xshift=2.3cm] (QM) {
    \textbf{Quantum mechanics}
};

\draw[arrow] (HJ.south) -- ++(-2.3cm,-0.45cm) -- (CL.north);
\draw[arrow] (HJ.south) -- ++( 2.3cm,-0.45cm) -- (QM.north);

\end{tikzpicture}

\caption{
Schematic summary of the framework introduced here.
A single structural theory---the HJS formulation---contains classical mechanics as the limit $|\kappa|\to 0$ and quantum mechanics as the regime where $\mathrm{Re}\,\kappa\neq 0$.  
Both emerge from the same underlying complexified HJ structure. 
\label{fig:parent}
}
\end{figure}

Taken together, these results identify the HJS
framework as a complementary formulation of classical mechanics, standing
alongside the Newtonian and Lagrange--Hamilton descriptions.
Its linear structure provides an alternative lens through which classical dynamics
may be represented, while its extended parameter space naturally accommodates the
mathematical ingredients characteristic of quantum theory.
In addition to its conceptual implications, the HJS representation clarifies why
wave-based and amplitude-inspired techniques have proved so effective in modern
post-Minkowskian and scattering approaches to gravitational dynamics: it furnishes
a purely classical linear representation in which such methods arise naturally.
The remainder of this work develops the structural consequences of the HJS system
and outlines how its linear formulation may serve as a unifying tool in both
classical and quantum contexts.
As a schematic summary of this viewpoint, Fig.~\ref{fig:parent} illustrates how
both classical and quantum mechanics emerge as distinct limits of the same HJS
structure.

Traditional routes from classical to quantum theory rely on intrinsically 
nonclassical inputs.  
Canonical quantization introduces operator commutators by fiat, 
path-integral quantization assumes a complex phase $e^{iS/\hbar}$, 
and semiclassical Wentzel--Kramers--Brillouin approaches insert a complex ansatz into an 
already-quantum equation.  
In all cases, linearity, superposition, and probabilistic interpretation 
are assumed rather than derived, and the connection to the underlying 
HJ structure remains indirect.

The analysis developed here shows that these quantum ingredients emerge 
automatically once the HJ ensemble is minimally complexified.  
Thus quantum mechanics appears not as a set of independent postulates, 
but as a particular regime of a broader classical-compatible structure. From this perspective, the familiar linear structure of quantum theory is
simply the most economical representation of classical HJ dynamics
once expressed in complex form.

\section*{Complex Embedding of HJ Dynamics}

The HJ formulation expresses classical dynamics in terms of a 
single phase function \(S(q,t)\).  
Its spatial gradient gives the canonical momentum,
\begin{equation}
p(q,t)=\nabla S(q,t),
\label{p_def}
\end{equation}
and therefore the associated velocity field
\begin{equation}
v(q,t)=\frac{\nabla S}{m}.
\label{v_def}
\end{equation}
The potential \(V(q,t)\) enters through the nonlinear first-order HJ equation
\begin{equation}
\partial_t S + \frac{(\nabla S)^2}{2m} + V = 0 .
\label{HJ}
\end{equation}
Once \(S(q,t)\) is known, the flow \eqref{v_def} determines the trajectories of 
all classical particles.

\subsection*{Ensemble evolution}

To describe an ensemble transported by this flow, introduce a density 
\(\rho(q,t)\) obeying the continuity equation
\begin{equation}
\partial_t \rho + \nabla\!\cdot\!\bigl(\rho\, v \bigr)=0 .
\label{continuity}
\end{equation}
Given an initial distribution \(\rho(q,0)\), the velocity field derived from 
\(S\) uniquely transports it through Eq.~\eqref{continuity}: the phase \(S\) 
specifies the flow, while \(\rho\) supplies the statistical weighting along it.

It is convenient to parameterize the nonnegative density as
\begin{equation}
\rho = R^2 ,
\label{rho_R2}
\end{equation}
with \(R(q,t)\) real.  
Substituting \eqref{rho_R2} into \eqref{continuity} yields an equivalent 
first-order equation for \(R\),
\begin{equation}
\partial_t R 
+ \frac{1}{m}\,\nabla R\!\cdot\!\nabla S
+ \frac{R}{2m}\,\nabla^2 S = 0 .
\label{R_eq}
\end{equation}
Together, Eqs.~\eqref{HJ} and \eqref{R_eq} form a closed evolution system for 
the pair \((R,S)\).

\subsection*{Motivation for a complex embedding}

The HJ equation \eqref{HJ} is nonlinear in the phase \(S\), with 
the kinetic term \((\nabla S)^2/2m\) supplying its essential nonlinearity.  
By contrast, many powerful evolution equations in theoretical physics are linear 
in their fundamental variables, admitting superposition, spectral methods, and 
wave-like representations that have no analogue in the HJ 
formulation.

This leads to a structural question of independent interest:
\begin{quote}
\emph{Can the coupled classical fields \((R,S)\) be re-expressed as a single 
complex field \(\psi\) satisfying a \emph{linear} evolution equation, while still 
reproducing the HJ and continuity dynamics exactly?}
\end{quote}
If such a representation exists, it would replace the nonlinear kinetic 
structure of Eq.~\eqref{HJ} with a linear differential operator acting on 
\(\psi\), providing a new functional viewpoint on classical mechanics and 
revealing potential connections to linear wave equations in other areas of 
physics.

\subsection*{General complex reparameterization}

We therefore introduce a complex field \(\psi(q,t)\) through the most general 
invertible decomposition
\begin{equation}
\psi = f(R,S)\, e^{i\, g(R,S)},
\label{general_ansatz}
\end{equation}
with \(f\) and \(g\) real functions.  
No quantum assumptions enter: \(\psi\) serves solely as an auxiliary 
representation of the classical variables.  
Local invertibility of \((R,S)\mapsto\psi\) ensures that no dynamical 
information is lost.

\subsection*{Structural requirements for a linear embedding}

To identify embeddings capable of unifying Eqs.~\eqref{HJ} and 
\eqref{R_eq} into a single equation for \(\psi\), we impose two minimal 
structural requirements motivated entirely by the desire for a \emph{linear} 
time-evolution equation that preserves the first-order character of the 
classical dynamics in \(\partial_t\).

\medskip
\noindent\textbf{(i) First-order time dependence.}
Both \eqref{HJ} and \eqref{R_eq} are first order in time.  
Any evolution equation for \(\psi\) reproducing them must therefore remain 
first order in \(\partial_t\psi\), even though it may contain higher-order 
spatial derivatives (as in the Schr\"odinger equation).
Reparameterizations that generate higher-order or mixed time derivatives when 
expressed in terms of \((R,S)\) are thus excluded.

\medskip
\noindent\textbf{(ii) Absence of nonlinear gradient structures.}
A generic map of the form \eqref{general_ansatz} produces nonlinear terms such 
as \((\nabla\psi/\psi)^2\) when Eqs.~\eqref{HJ} and \eqref{R_eq} are 
rewritten in terms of \(\psi\).  
Such contributions cannot be absorbed into any linear differential operator 
acting on \(\psi\) and must therefore vanish identically.

\medskip
These two requirements are extremely mild:  
they do \emph{not} constrain the eventual linear operator acting on \(\psi\),  
they do \emph{not} presuppose quantum structure,  
and they do \emph{not} introduce new physics.  
They simply encode the minimal compatibility conditions that any linear complex 
embedding of the HJ system must satisfy.

\subsection*{Uniqueness of the allowed complex map}

Imposing Requirement (i) restricts the \(S\)-dependence of \(f\) and \(g\).
When the time derivative \(\partial_t\psi\) is expressed through the variables 
\((R,S)\), both \(\partial_t S\) and \(\partial_t R\) appear, 
with \(\partial_t R\) fixed by the ensemble continuity equation. 
For the resulting evolution equation for \(\psi\) to remain linear and first order 
in \(\partial_t\psi\), the prefactor multiplying \(\partial_t S\) must reduce to a 
fixed complex constant---otherwise the evolution equation would acquire an explicit 
\((R,S)\)-dependent coefficient and could not be linear.
This requirement forces
\[
\partial_S g = c_1, \qquad \partial_S f = c_2\, f ,
\]
and hence
\[
g(R,S)= c_1 S + A(R), \qquad f(R,S)= B(R)e^{c_2 S}.
\]
with \(A(R)\) and \(B(R)\) arbitrary real functions.

Requirement (ii) eliminates the nonlinear gradient structure 
\((\nabla\psi)^2/\psi^2\).  
Substituting the above forms into the transformed HJ equation 
shows that all such nonlinearities cancel iff
\[
A'(R)=0,
\qquad
R B'(R)-B(R)=0.
\]
Thus
\[
A(R)=c_3,
\qquad
B(R)= R\, e^{c_4}.
\]
Removing an irrelevant global phase and normalization yields the unique 
admissible embedding
\begin{equation}
\label{psi_final}
\psi = R\, e^{i S/\kappa},
\qquad
\kappa = \frac{1}{c_1 - i c_2}.
\end{equation}

The full technical derivation of this uniqueness statement is presented in the
 Supplemental Material.
\section*{The HJS Equation}

Given the unique embedding 
\eqref{psi_final}, 
the phase field may be reconstructed as 
\begin{equation}
S = -\,i\kappa\,\ln(\psi/R).
\label{S_from_psi_sec23}
\end{equation}
Substituting \eqref{S_from_psi_sec23} into both the continuity equation 
\eqref{R_eq} and the HJ equation \eqref{HJ}, and eliminating 
$\partial_t R$ using \eqref{R_eq}, all phase--dependent nonlinearities cancel 
exactly. The HJ equation reorganizes into
\begin{align}
\partial_t S + \frac{(\nabla S)^2}{2m} + V
&=
\frac{1}{\psi}\!\left(
-i\kappa\,\partial_t\psi
-\frac{\kappa^{2}}{2m}\,\nabla^{2}\psi
+V\psi
\right)
\nonumber\\[1mm]
&\qquad
+\,\frac{\kappa^{2}}{2m}\frac{\nabla^{2}R}{R}.
\label{HJ_reorganize_sec23}
\end{align}

The only surviving correction is proportional to 
$\kappa^2$
and depends solely on 
$R$. Defining
\begin{equation}
Q[R] \equiv -\,\frac{\kappa^{2}}{2m}\,\frac{\nabla^{2}R}{R},
\label{Q_def_sec23}
\end{equation}
and moving it to the left-hand side produces the modified HJ equation
\begin{equation}
\partial_t S 
+ \frac{(\nabla S)^2}{2m}
+ V + Q[R] = 0.
\label{HJ_modified_sec23}
\end{equation}

With this identification, the right-hand side of 
\eqref{HJ_reorganize_sec23} collapses to a purely \emph{linear} expression in
$\psi$, giving the Schr\"odinger-like equation
\begin{equation}
\label{Sch_sec23}
i\kappa\,\partial_t\psi
=
-\frac{\kappa^{2}}{2m}\,\nabla^{2}\psi
+V\psi.
\end{equation}
Equations \eqref{HJ_modified_sec23} and \eqref{R_eq} are therefore 
\emph{identically equivalent} to the single complex linear equation
\eqref{Sch_sec23} on any region where the polar variables \((R,S)\)
provide a regular chart for the map \(\psi = R e^{iS/\kappa}\).\footnote{That is, on connected regions where \(R\neq 0\) and a continuous branch of \(S\) can be chosen. At nodal sets or across phase-branch changes, the variables \((R,S)\) may fail to furnish a single regular parametrization, although the linear equation for \(\psi\) itself remains well defined.}

\medskip
\noindent\textbf{Classical limit.}  
Since \(Q[R]\propto \kappa^{2}\), it vanishes as \(|\kappa|\to 0\) wherever the polar variables remain regular and \(R\) does not develop \(\kappa\)-dependent singular structure.\footnote{Equivalently, this classical limit is understood locally in regimes for which \(\kappa^{2}\nabla^{2}R/R \to 0\). Near turning points, caustics, or nodal sets, the single-phase polar decomposition may cease to define a regular local classical branch, and a uniform wave description is generally more appropriate.}
In this limit, \eqref{HJ_modified_sec23} reduces to the HJ 
equation \eqref{HJ}, while \eqref{R_eq} remains unchanged.  
Thus the nonlinear HJ system is recovered \emph{exactly} as the 
\(|\kappa|\to 0\) limit of the linear equation \eqref{Sch_sec23}.

\medskip
\noindent 
Equations \eqref{Sch_sec23} and 
\eqref{HJ_modified_sec23}--\eqref{R_eq} therefore provide two equivalent 
formulations of the same underlying dynamics, expressed respectively in linear complex and nonlinear polar variables.  
We refer to this unified structure as the 
\emph{HJS system}.

\begin{figure}[t]
\centering
\begin{tikzpicture}[
    node distance=1.0cm,
    box/.style={draw, thick, rounded corners,
        align=center, minimum width=5.0cm,
        inner sep=6pt},
    arr/.style={-stealth, thick}
]

\node[box] (A) {
\textbf{Classical HJ structure}\\[2pt]
$R^2=\rho$, \quad $S$ solves HJ, \\
continuity equation
};

\node[box, below=1.2cm of A] (B) {
\textbf{Minimal and unique complex embedding}\\[2pt]
$\psi = R\, e^{iS/\kappa}$\\
Linear HJS equation\\
Born probability from structural consistency
};

\node[box, below=1.2cm of B] (C) {
\textbf{Emergent quantum regime}\\[2pt]
Superposition \\
Operators and commutators \\
Unitary evolution ($\theta=0$) 
\\
Uncertainty principle
};

\draw[arr] (A) -- (B);
\draw[arr] (B) -- (C);

\end{tikzpicture}

\caption{
Structural pipeline of the HJS formulation.
Classical HJ dynamics and ensemble continuity admit a unique
complex embedding.  
This minimal complexification yields the linear HJS/Schr\"odinger equation,
from which standard quantum structures---superposition, operators,
commutators, unitary evolution, and uncertainty---emerge as consistency
requirements.
\label{fig:flow}}
\end{figure}
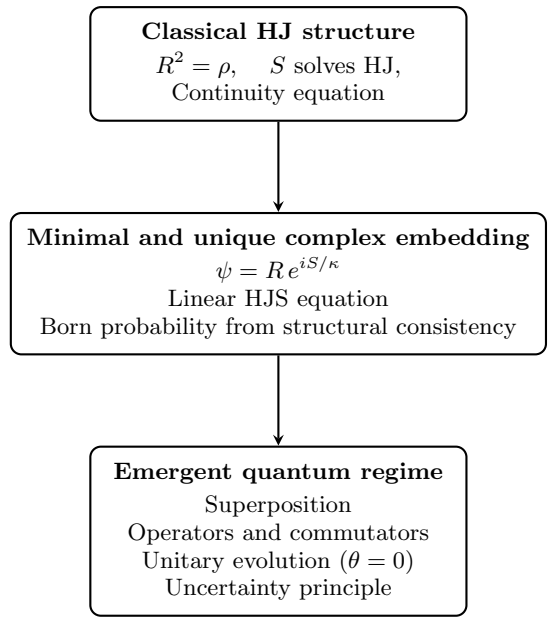

The construction above reveals a simple but powerful flow:
a purely classical HJ ensemble,
together with the uniquely admissible complex embedding,
collapses into a single linear equation whose structure
automatically contains the ingredients usually associated
with quantum theory.
For clarity, this structural pipeline is summarized in
Fig.~\ref{fig:flow}.

Historically, the Schr\"odinger equation was discovered through analogies with
the HJ formulation of classical mechanics \cite{Schrodinger1926}.
Despite this close conceptual link, the precise structural route by which a
classical HJ ensemble reorganizes into a linear wave equation
has remained implicit.
The present construction makes this connection explicit within a single,
self-consistent framework.

\medskip
\noindent\emph{Relation to the quantum potential.}
The functional form of the correction term \(Q[R]\) coincides with the
quantum potential originally appearing in the hydrodynamic reformulation
of quantum mechanics by Madelung \cite{Madelung1927}, 
and later identified and interpreted as a distinct dynamical potential by Bohm
\cite{Bohm1952}.
In the present framework, however, \(Q[R]\) is not introduced by starting
from the Schr\"odinger equation and rewriting it in polar form. Rather,
it appears together with the Schr\"odinger-type equation itself as a
direct consequence of the unique admissible linear complex embedding of
the classical HJ ensemble.

\section*{Emergent quantum structures}

Once the HJ system is embedded into the linear equation   \eqref{Sch_sec23}, 
a number of hallmark quantum structures arise automatically.
None of them are imposed; they follow solely from the linearity of the
Schr\"odinger--like evolution and the representation 
$\psi = R e^{iS/\kappa}$.

\subsection*{Linearity and complex vector space}

Equation \eqref{Sch_sec23} is linear in $\psi$.
Therefore every rescaling $C\psi$ and every superposition
\[
C_1\psi_1 + C_2\psi_2
\]
again solves the same equation.
Thus the minimal complex embedding endows the classical HJ 
framework with a complex vector space of states, a structure absent in the
original $(R,S)$ formulation.
This linear structure holds for any nonzero complex $\kappa$; no assumption
on its reality is required at this stage.

\subsection*{Operators and eigenvalue equations}

Classically, the momentum field is
\[
p = \nabla S .
\]
Using $\psi = R e^{iS/\kappa}$, this becomes
\begin{align}
p
&= \nabla S
 = - \,i\kappa\,\frac{\nabla\psi}{\psi}
   +i \kappa\,\frac{\nabla R}{R}.
\label{p_split_latex}
\end{align}
The first term is linear in $\psi$; the second is intrinsically nonlinear.
Demanding that the momentum operation act linearly on solutions of
\eqref{Sch_sec23} forces us to isolate the linear part:
\[
\hat p\,\psi \equiv -\,i\kappa\,\nabla\psi .
\]
This definition guarantees compatibility with the linear evolution and becomes
exact for momentum eigenstates, whose modulus must be spatially uniform
($\nabla R = 0$).
Thus the momentum operator emerges uniquely from the consistency of the
complex embedding:
\[
\hat p = -\,i\kappa\,\nabla .
\]

\subsection*{Canonical commutators and uncertainty}

With $\hat q = q$, the standard computation yields, for any smooth $\psi$,
\begin{align}
(\hat q \hat p - \hat p \hat q)\psi
= i\kappa\,\psi .
\end{align}
Hence the canonical commutator
\[
[\,\hat q,\hat p\,] = i\kappa
\]
is not postulated but required by linearity together with the definition of
$\psi$.
We postpone any probabilistic interpretation of this commutator until the
Born rule is established in the next section.

\section*{Emergence of the Born Rule and the Role of \texorpdfstring{$\kappa$}{kappa}}

Any probability assignment $P(\psi)$ constructed from the complexified field
$\psi = R e^{iS/\kappa}$ must ultimately reproduce the classical ensemble
density $\rho$.
The statement $P(\psi)=\rho$ appears simple, but it is in fact fixed by three
purely \emph{classical} consistency requirements.

\medskip
\noindent\textbf{(i) Recovery of the classical ensemble.}
The complex field $\psi$ is introduced only as a representation of the
underlying HJ ensemble variables $(R,S)$.
Therefore any legitimate probability assignment must reduce to the original
classical density when the deformation is removed:
\begin{equation}
\lim_{|\kappa|\to 0} P(\psi;\kappa)=\rho.
\label{P_classical_limit}
\end{equation}

\medskip
\noindent\textbf{(ii) Independence from the HJ phase $S$.}
The HJ phase $S(q,t)$ encodes dynamical information, such as momentum flow and
energy, whereas $\rho(q,t)$ is a purely statistical quantity specifying the
ensemble weight.
A probability assignment intended to reproduce the ensemble density therefore
cannot depend on $S$ at fixed $\rho$ even when $\kappa\neq 0$.
Equivalently,
\begin{equation}
\frac{\partial P}{\partial S}=0,
\qquad\Rightarrow\qquad
P(\psi;\kappa)=\mathcal{P}(\rho;\kappa).
\label{P_free_of_S}
\end{equation}

\medskip
\noindent\textbf{(iii) Independence from the deformation scale $|\kappa|$.}
The deformation parameter $\kappa$ carries the same dimension as $S$,
namely that of action (or angular momentum).
Once the probability assignment is required to be independent of $S$,
there is no natural way for $|\kappa|$ to enter a classical probability
density on its own.
Moreover, the notion of probability itself remains classical, even though
the dynamics is now $\kappa$-deformed.
Classical probabilities therefore carry no intrinsic scale and cannot depend on
the magnitude of the deformation parameter:
\begin{equation}
\frac{\partial P(\psi;\kappa)}{\partial|\kappa|}=0.
\label{P_free_of_kappa}
\end{equation}

\medskip
Taken together, Eqs.~\eqref{P_classical_limit}, \eqref{P_free_of_S}, and
\eqref{P_free_of_kappa} leave only one possibility:
\begin{equation}
\label{P_rho}
P(\psi)=\rho.
\end{equation}

Writing $\psi = R e^{iS/\kappa}$, the result~\eqref{P_rho} allows the general
representation
\begin{equation}
P(\psi)=R^{2}
=
\psi^{\star\,(1-i\theta)}
\,
\psi^{\,1+i\theta},
\qquad
\theta \equiv \frac{\mathrm{Im}\,\kappa}{\mathrm{Re}\,\kappa}.
\label{general_prob_kappa}
\end{equation}
Thus, even before any symmetry input, the Born rule emerges as the unique
classical-compatible probability assignment.

\subsection*{$\kappa$--pseudo unitarity}

For a general complex deformation parameter $\kappa$, the $\kappa$--Born
probability induces a natural $\theta$--dependent inner product,
\begin{equation}
\langle \phi, \psi \rangle_\theta
:= \int dq\;
\phi^{\star(1-i\theta)}\,
\psi^{\,1+i\theta} .
\label{inner_product_theta}
\end{equation}
The HJS equation is linear, so any superposition of solutions is again a
solution, while the continuity equation ensures conservation of
$\langle\psi,\psi\rangle_\theta$ for all evolved states.

Let $\psi(t)=U(t)\psi(0)$.  
Demanding conservation of the $\theta$--inner product for all pairs
$\phi,\psi$,
\begin{equation}
\langle U(t)\phi,U(t)\psi\rangle_\theta
+ \langle U(t)\psi,U(t)\phi\rangle_\theta
=
\langle \phi,\psi\rangle_\theta
+ \langle \psi,\phi\rangle_\theta ,
\end{equation}
is \emph{equivalent} to the $\kappa$--pseudo unitarity condition
\begin{equation}
{U^\dagger(t)}^{\,1-i\theta}\;
U(t)^{\,1+i\theta}
= \mathbf 1 .
\label{kappa_pseudo_unitarity}
\end{equation}
This is the natural $\kappa$--generalization of standard unitarity.  
In the remainder of this work we focus on the physically relevant
time-reversal-symmetric case $\theta=0$, where pseudo unitarity reduces to the
ordinary unitary condition.

\subsection*{Time-reversal symmetry and the reality of $\kappa$}

The HJS formalism itself allows an arbitrary complex deformation parameter
$\kappa$.
However, if one requires the deformed dynamics to inherit the
time-reversal symmetry of the classical HJ system,
\[
t\to -t,\qquad S\to -S,\qquad R\to R ,
\]
then the corresponding transformation of the complex field,
\[
\psi(t)=R e^{iS/\kappa}
\;\longrightarrow\;
\psi_{\rm TR}(t)=R e^{-iS/\kappa},
\]
must again define a solution of the same evolution equation.
This requirement selects a real deformation parameter,
\begin{equation}
\mathrm{Im}\,\kappa = 0 .
\end{equation}

This conclusion can also be seen directly from the 
modified HJ plus continuity formulation.  
When $\kappa$ is complex, the deformation term  
\[
Q[R] \;\propto\; \kappa^{2}\,\frac{\nabla^{2}R}{R}
\]
acquires an imaginary part.  
The resulting equations for $(R,S)$ no longer transform into themselves under
$t\to -t$, and the continuity equation develops a source-like term when written
in terms of the effective velocity field 
$\nabla S/m$.  
In other words, ${\rm Im}\,\kappa\neq0$ introduces a structural
time-asymmetry on the classical side of the HJS system, corresponding to a
dissipative or anti-dissipative flow of the ensemble density.

Only when $\kappa$ is real does the deformation preserve the 
$(R,S)$--equations under time reversal, and the dynamics remains
non-dissipative in the ensemble sense.  
In this case, Eq.~\eqref{general_prob_kappa} reduces to the quadratic Born rule,
\begin{align}
P(\psi)=R^{2}=|\psi|^{2},
\end{align}
and the familiar time-reversal symmetry of standard quantum mechanics is
recovered.

\subsection*{Heisenberg uncertainty and ensemble uplift}

Once the Born rule supplies a probabilistic interpretation to $|\psi|^{2}$, the
canonical commutator $[\hat q,\hat p]=i\kappa$ immediately yields
\[
\Delta q\,\Delta p \ge \frac{|\kappa|}{2}.
\]
Thus the uncertainty principle is a structural consequence of the complex 
embedding.

Classically, $\rho$ describes an ensemble of systems.  
For finite $\kappa$, the nonlinear coupling between $(R,S)$ prevents $\rho$ 
from collapsing to a delta distribution, even for a single system.  
The $\kappa$--deformed theory therefore elevates the ensemble picture to a 
fundamental feature.

\subsection*{Relation to Gleason-type results}

In standard quantum theory, Gleason-type theorems imply that any non-contextual
probability measure must be quadratic \cite{Gleason1957}.  
Here the quadratic structure arises much earlier---directly from classical
consistency at the point where the complex embedding is introduced.  
Gleason's theorem then acts as a consistency check rather than the origin of
the Born rule.

\section*{Representations and Physical Interpretation}

The HJS system may be expressed in two equivalent languages:
(i) the pair $(R,S)$ obeying the deformed HJ equation together
with the continuity equation, and
(ii) the single complex waveform $\psi = R\,e^{iS/\kappa}$ obeying a linear
Schr\"odinger--type equation.
The two descriptions encode the same physics; they differ only in
convenience and perspective.

\subsection*{The $(R,S)$ representation}

In the $(R,S)$ formulation, the meaning of the variables remains entirely
classical.
The field $R^{2}$ represents the ensemble density, $\nabla S$ defines the
momentum flow, and the combination $-\partial_t S - Q[R]$ plays the role of
an effective Hamiltonian.
The dynamical content of the theory resides completely in the coupled
evolution of $(R,S)$ through the HJS equations.

Nothing prevents one from solving the dynamics directly in terms of these
variables.
Given initial data $(R(q,0),S(q,0))$, the subsequent evolution
$(R(q,t),S(q,t))$ is fixed uniquely and deterministically for all times.
No stochastic ingredient or probabilistic postulate enters the evolution
itself; probability arises only at the level of ensemble interpretation and
observable extraction.

From this viewpoint, the distinction between classical and quantum mechanics
is structurally modest.
It is governed by a single deformation parameter $\kappa$:
for $|\kappa|\to0$ the system reduces to classical HJ theory,
while for $\mathrm{Re}\,\kappa\neq0$ additional algebraic structures become
available.
What changes is not the physical meaning of $R$ and $S$, but the way their
evolution is represented.

\subsection*{The $\psi$ representation}

The complex waveform $\psi$ packages the same real ensemble variables $(R,S)$
into a single object whose evolution is linear whenever
$\mathrm{Re}\,\kappa\neq0$.
This linearization makes available a powerful set of algebraic tools,
including superposition, spectral decompositions, eigenfunctions,
commutators, and operator calculus, which greatly simplify the analysis of
the dynamics.

From this perspective, the operator formalism is not an additional physical
postulate but a convenient linear calculus acting on $\psi$.
Observable quantities may be computed either from the linear evolution of
$\psi$ or directly from the ensemble variables $(R,S)$ using $R^{2}$ and
$\nabla S$.
Both procedures are exactly equivalent.

\subsection*{Unified viewpoint}

The HJS framework therefore places classical and quantum descriptions on the
same footing.
The complex waveform $\psi$ serves as a unifying representation that renders
the dynamics linear, while $R$ and $S$ remain the directly physical fields
describing the ensemble.
Quantum structures emerge not from new postulates, but from the linear
organization of an underlying classical-compatible ensemble dynamics.
This equivalence is most naturally formulated at the level of ensembles,
rather than individual trajectories. 

A benchmark example in the Supplemental Material illustrates how the ensemble evolution is organized within this framework.
The extension to internal degrees of freedom, including spin, is discussed in a separate section of the Supplemental Material, where the corresponding deterministic mechanism is outlined.

\subsection*{Eulerian ensemble vs Lagrangian ensemble}

Before proceeding further, it is worth emphasizing that the present
formulation implicitly adopts an Eulerian ensemble description, in which
the fields $(R,S)$ evolve as smooth functions over configuration space and
time.
At the classical level, this description is equivalent to a Lagrangian
ensemble of trajectories and admits the standard correspondence between
density evolution and characteristic flows.
For finite $\kappa$, the precise relation between these viewpoints may
involve additional subtleties that lie beyond the scope of the present
work.
Throughout this paper, we focus on the Eulerian ensemble formulation,
within which the HJS dynamics is fully deterministic for all values of
$\kappa$.

\subsection*{Remark on superfluid hydrodynamics}

Closely related Eulerian structures have long appeared in superfluid
hydrodynamics, where the velocity field is written in HJ form
as $v=\nabla S/m$.
Historically, however, this formulation has almost universally been
interpreted as an \emph{effective} description descending from an
underlying quantum many-body wavefunction.
From the viewpoint developed here, this ordering is conceptually reversed:
an Eulerian ensemble governed by $(R,S)$ already carries the minimal
structural data required for a linear complex description, while familiar
quantum structures emerge only at a later representational stage.

\begin{table*}[t]
\centering
\renewcommand{\arraystretch}{1.25}

\begin{tabular}{c c c c}
\hline
\textbf{Structure} &
\textbf{Classical HJ} &
\textbf{HJS formulation} &
\textbf{Quantum theory} \\
\hline

State object &
$(R,S)$ &
$\psi = R e^{iS/\kappa}$ &
$\psi$ \\
\hline

Evolution &
Nonlinear PDEs &
Linear HJS equation &
Linear Schr\"odinger equation \\
\hline

Probability &
$\rho = R^{2}$ &
Born rule from consistency &
Born rule postulated \\
\hline

Superposition &
Absent &
Follows from linearity &
Fundamental principle \\
\hline

Operators / algebra &
None &
Canonical commutators emerge &
Operator postulates \\
\hline

Uncertainty &
Absent &
$\Delta q\, \Delta p \ge |\kappa|/2$ &
Heisenberg principle \\
\hline

Time reversal &
Invariant &
Selects $\mathrm{Im}\,\kappa=0$ &
Requires Hermiticity \\
\hline

Classical limit &
--- &
$|\kappa| \to 0$ reproduces HJ &
$\hbar \to 0$ semiclassical \\
\hline
\end{tabular}

\caption{
Comparison of structural elements in classical HJ theory, the HJS formulation, and quantum mechanics for nonrelativistic single-particle systems.
}
\label{tab:comparison}
\end{table*}

\section*{Discussion and Outlook}

The HJS framework developed here exposes a 
remarkably tight structural bridge between classical and quantum mechanics.  
What begins as a minimal complexification of the HJ ensemble 
automatically generates mathematical features normally regarded as 
distinctively quantum: linear superposition, operator algebras, commutators, 
the Heisenberg uncertainty relation, and---when $\kappa$ is real---unitary 
time evolution.  
As emphasized throughout this work, the linear Schr\"odinger-type equation is
best understood as a computational representation of the underlying
HJ ensemble: it greatly simplifies solving classical evolution
problems, while the variables $(R,S)$ retain their direct physical meaning
even in regimes that are usually described as quantum.
In this perspective, quantum mechanics is not an independently postulated 
theory, but a particular \emph{regime} of a broader classical-compatible 
structure.  
To highlight the scope of this structural unification, 
Table~\ref{tab:comparison} summarizes the correspondences between 
classical HJ theory, the HJS formulation, and standard quantum 
mechanics.

Several conceptual and physical directions follow naturally.

\subsection*{Applications to scattering amplitudes and classical dynamics}
The linearization uncovered by the HJS map provides a structural
perspective on why wave-based and amplitude-inspired techniques have
proved remarkably effective in problems that are, at their core,
classically nonlinear.  
By rewriting HJ evolution as a linear operator equation for a complex field, the HJS framework makes
available the entire toolbox of Fourier methods, spectral
decompositions, and superposition techniques---without invoking any
quantization principle.

A particularly clear illustration appears in the post-Minkowskian
two-body problem, where modern scattering-amplitude methods have led to
substantial analytic progress, including recent high-precision and
geometric insights \cite{Goldberger:2004jt,Bern:2019nnu,Driesse:2024feo}.  
Within the HJS viewpoint, the effectiveness of these amplitude methods is
no longer surprising: once the classical dynamics is expressed in terms of
complexified HJ variables, the resulting linear structure
naturally accommodates the same tools familiar from quantum scattering  theory,
including basis decompositions and momentum-space representations.

Together with the independent advances made in self-force theory,
black-hole perturbation theory and quasi-normal modes, the Effective One
Body framework, and post-Newtonian radiation-flux methods \cite{Damour2016PMEOB},
these developments have significantly improved the analytic understanding of
gravitational dynamics across complementary regimes.

More broadly, the perspective developed here suggests that HJS-type
linearizations may benefit a wide range of classical systems whenever
nonlinear evolution can be reorganized into a linear flow.  
Reducing classical dynamics to a wave equation preserves interpretability,
complements numerical approaches by reducing reliance on black-box
procedures, and often reveals structural connections between otherwise
disparate areas of physics.

The linearization trades a nonlinear classical problem for a representation
involving infinite-mode expansions when $|\kappa|\to 0$.  
Such divergences are structural and well controlled---much like UV divergences in
field theory---and the analytic advantages of the linear formulation
overwhelmingly outweigh this cost.  
Recognizing this tradeoff clarifies why linear representations can uncover
features of classical evolution that remain opaque in purely nonlinear
formulations.

From this viewpoint, even long-familiar classical problems may contain
underexplored insights: had wave-based representations been applied to
certain systems earlier in history, some of the linear structures now
associated with quantum mechanics might have been recognized much sooner.

\subsection*{Toward a relativistic and field-theoretic generalization}

The HJS construction is not restricted to the nonrelativistic point-particle
setting.
Any classical system admitting an HJ formulation carries the same
structural data $(R,S)$ at the level of ensemble dynamics, and can therefore be
locally embedded into a complex field of the form
$
\psi = R\,e^{iS/\kappa}.
$
This observation extends the HJS mechanism to $N$-particle ensembles,
relativistic worldlines, and, at a formal level, to classical field theories
whose HJ description is expressed in functional form.

As a representative example, consider a relativistic scalar field $\phi(\mathbf{x},t)$. Its classical HJ equation may be written as
\begin{align} \partial_t S[\phi,t] + \int d^3x \Big[ \frac{1}{2} & \left( \frac{\delta S}{\delta\phi(\mathbf{x})} \right)^2 \nonumber \\ & + \frac{1}{2}(\nabla\phi)^2 + \frac{1}{2}m^2\phi^2 + V(\phi) \Big] = 0 . \nonumber \end{align}
At the level of ensemble dynamics, this functional equation shares the same
structural ingredients as the nonrelativistic point-particle case, now
formulated on an infinite-dimensional configuration space.
This observation suggests that a minimal complex embedding of the form
$\psi[\phi,t] = R[\phi,t]\,e^{iS[\phi,t]/\kappa}$
may be introduced in close analogy with the finite-dimensional setting.

While a detailed analysis is beyond the scope of the present work, the formal
similarity indicates that the coupled HJ and continuity equations
may admit a linear Schr\"odinger-type reorganization at the ensemble level,
as illustrated in the scalar-field case in~\cite{Zhang:2026pde}.
Similar considerations are expected to apply to more general classical field
configurations admitting an HJ formulation.
If the amplitude $R$ carries additional internal indices, the same construction
naturally accommodates representation-valued wavefunctionals
$\psi^{A}=R^{A}e^{iS/\kappa}$, hinting at a possible extension to systems with
spin or other internal structure.

General relativity also admits an HJ formulation
\cite{DeWitt1967,Kuchar:1991qf}; however, its constraint structure and
diffeomorphism symmetry render any direct extension of the present analysis
nontrivial.
Nevertheless, the minimal complexification mechanism isolates which elements of
quantum structure arise from the most primitive features of classical ensemble
flow, offering potential guidance in regimes---such as black-hole interiors,
cosmological singularities, or strongly curved backgrounds---where standard
quantization procedures are difficult to apply.

Viewed broadly, the HJS map provides a unified procedure that associates a linear
complex representation to any classical system with an HJ
description.
Standard quantum mechanics corresponds to the special case in which unitarity
and time-reversal symmetry select a real deformation parameter $\kappa$, while
relaxing these conditions leaves open a larger family of structurally
self-consistent quantum-like extensions.

\subsection*{The ratio $\theta={\rm Im}\,\kappa/{\rm Re}\,\kappa$ as a structural time-asymmetry parameter}

A final structural aspect of the HJS embedding concerns the ratio
$\theta={\rm Im}\,\kappa/{\rm Re}\,\kappa$.
Although unnecessary for the emergence of standard quantum mechanics,
$\theta$ parametrizes the departure from the real-$\kappa$ branch.
For $\theta\neq 0$, the HJS system remains formally self-consistent, but no
longer preserves time-reversal symmetry.
At the structural level, unified linear HJS evolution still requires a common
deformation parameter $\kappa$ across interacting sectors, including both its
magnitude and its phase; otherwise the dynamics can no longer be assembled into
a single universal complex evolution law.
In this sense, the universality of $\kappa$ mirrors that of Planck's constant,
not as an independent postulate, but as a structural requirement for unified
linear dynamics.

At the phenomenological level, a nonzero $\theta$ may in principle generate
extremely weak scale-dependent effects, since the corresponding deformation can
be associated with an effective complexification of the energy eigenvalue
structure.
This opens the possibility of very slow amplification or decay in time
evolution over long time scales, although any such effect is likely to be so
small that it may remain invisible even over cosmological durations.
The same deformation also modifies the complex probability structure and
therefore affects phase-sensitive interference processes already at leading
order. For example, a two-path pattern of the form
\[
I(\mathbf{x}) \,\propto\,
\bigl|\,\psi_1(\mathbf{x}) + \psi_2(\mathbf{x})\,\bigr|^{2}
\;\longrightarrow\;
I(\mathbf{x})\,[\,1 + \theta\,F(\mathbf{x}) + O(\theta^{2})\,],
\]
acquires a characteristic $\theta$-dependent correction.

More broadly, because both the generalized Born rule and the underlying complex
phase structure are modified when $\theta\neq0$, such a deformation may in
principle influence coherent phenomena quite generally, including
Aharonov--Bohm-type phases, entanglement-related quantities such as entanglement
entropy, and, at a more formal level, aspects of field-theoretic symmetry
structure such as CPT symmetry.
At the same time, all such effects are likely to be extremely small in realistic
regimes, making direct detection very difficult; cosmological evolution may
already impose strong constraints on any appreciable departure from
$\theta=0$.
For this reason, in the present work we regard the $\theta\neq0$ branch mainly
as a formally allowed extension of the real-$\kappa$ framework, while a more
detailed phenomenological analysis is left for future work.

\section*{Acknowledgements}
Y.Z. is grateful to Rong-Gen Cai, Bin Chen, Lin Chen, Yun Fang, Song He (ITP, CAS),
Song He (Ningbo U.), Sen-Yue Lou, Haiyang Yan, Su Yi, Chi Zhang,
and Huaxing Zhu for useful discussions.
This work is supported by the National Natural Science Foundation of China
under Grant No.~12405086.

\bibliographystyle{physics}

\bibliography{Refs}

\widetext

\clearpage

\onecolumngrid

\begin{center}
{\large\bfseries SUPPLEMENTARY INFORMATION}
\end{center}

\vspace{1.5em}


\section{Uniqueness of the Minimal Complex Embedding}

This section provides the technical derivation supporting the uniqueness
statement in the main text: among all complex maps 
$\psi=f(R,S)e^{i g(R,S)}$ from the HJ ensemble variables
$(R,S)$ to a single complex wavefunction, only
$
\psi = R\,e^{iS/\kappa}
$
yields a Schr\"odinger-type linear evolution for $\psi$ while remaining exactly
equivalent to the HJ and continuity equations.

\subsection*{Linearity constraint}
Using the continuity equation to eliminate $\partial_t R$, the time derivative of
the general ansatz becomes
\begin{align}
\frac{\partial_t\psi}{\psi}
=&
\Bigl(\frac{\partial_S f}{f}+ i\,\partial_S g\Bigr)\partial_t S
-
\Bigl(\frac{\partial_R f}{f}+ i\,\partial_R g\Bigr)
\left(
\frac{\nabla R\cdot\nabla S}{m}
+ \frac{R}{2m}\nabla^2 S
\right).
\label{eq:A1}
\end{align}

For the evolution equation of $\psi$ to remain linear in $\partial_t\psi$,  
the coefficient multiplying $\partial_t S$ must be a constant.
This requires
$\partial_S g=c_1$ and $\partial_S f=c_2 f$,
with $c_1,c_2$ real.
Thus
\begin{align}
    \label{eq:gf_form}
    g(R,S)=c_1 S + A(R),
\qquad
f(R,S)=B(R)e^{c_2 S}.
\end{align}

Introducing the deformation parameter $\kappa = 1/(c_1 - i c_2)$,
substituting \eqref{eq:gf_form} into \eqref{eq:A1},
and then inserting \eqref{eq:A1} into the HJ equation yields
\begin{align}
\label{HJinter}
&  \partial_t S 
+ \frac{(\nabla S)^2}{2m}
+ V
= -\frac{i \kappa \partial_t \psi }{\psi }
+ \frac{(\nabla S)^2}{2 m}+V
+
\frac{\kappa  (R \nabla^2 S +2  \nabla  R \nabla S)
\left( A'(R)B(R) -i B'(R)\right)}{2  m B(R)}
\,.
\end{align} 

\subsection*{Cancellation of nonlinear gradient terms}

Substituting \eqref{eq:gf_form} into the ansatz gives
\[
\psi = B(R)\,e^{\,i(S/\kappa + A(R))}.
\]

Solving this relation for $S$ in terms of $(\psi,R)$ yields expressions for 
$\nabla S$ and $\nabla^2 S$.
Inserting these into the reorganized HJ relation
\eqref{HJinter} shows that all nonlinear gradient structures combine into a
single term of the form $(\nabla\psi)^2/\psi^2$, with an overall coefficient
\[
i\,B(R)\,R\,A'(R) + R B'(R) - B(R).
\]
Requiring the absence of such nonlinear terms therefore gives
$A'(R)=0$,
 and 
$R B'(R)-B(R)=0$.
The resulting functions are
\[
A(R)=c_3,
\qquad
B(R)= R\,e^{c_4},
\]
with real constants $c_3,c_4$.
Discarding the irrelevant overall factor $e^{c_4+i c_3}$ leads to the unique
admissible embedding,
\[
\boxed{\psi = R\,e^{iS/\kappa}}.
\]

\section{Harmonic oscillator as a benchmark example}

In this section we illustrate the general framework developed in the
main text using the one-dimensional harmonic oscillator.
All results collected here are standard and well known.
They are included solely to provide a transparent benchmark and to
visualize the abstract structures discussed in the main text in a
concrete setting.
No new calculations are involved.
For simplicity, we restrict attention throughout this section to real
values of the deformation parameter $\kappa$.

\subsection*{Model and equivalent formulations}

We consider a one-dimensional harmonic oscillator with Hamiltonian
\begin{equation}
H(q,p)=\frac{p^2}{2m}+\frac12 m\omega^2 q^2,
\end{equation}
and set $m=\omega=1$ for notational simplicity.

The dynamics may be described equivalently in two languages:
\begin{itemize}
\item[(i)] the Schr\"odinger equation
\begin{equation}
i\kappa\,\partial_t\psi
=
\left(
-\frac{\kappa^2}{2}\partial_q^2+\frac12 q^2
\right)\psi,
\label{HO_Sch}
\end{equation}
\item[(ii)] the amplitude--phase decomposition
\begin{equation}
\psi(q,t)=R(q,t)\,e^{iS(q,t)/\kappa},
\end{equation}
which yields the HJ equation with quantum potential
together with the continuity equation for $\rho=R^2$.
\end{itemize}
Both formulations are exactly equivalent and describe the same
underlying dynamics.

\subsection*{Initial data}

We choose smooth initial data in the form of a polynomially modified Gaussian wave packet,
\begin{equation}
\label{initialRq0}
R(q,0)=(q^2+\varepsilon^2)\,
\exp\!\left(-\frac{q^2}{2\sigma^2}\right),
\qquad
S(q,0)=p_0\,q ,
\end{equation}
with fixed parameters $\varepsilon$, $\sigma$, and $p_0$.
The linear phase $S(q,0)=p_0 q$ corresponds to a uniform initial momentum
shift of the ensemble and implies
\begin{equation}
\langle p\rangle(0)=p_0 .
\end{equation}

\subsection*{Expectation values and fluctuations}

For the harmonic oscillator, expectation values and second moments obey
closed equations of motion.
In particular, the expectation values follow the classical trajectories
exactly,
\begin{equation}
\langle q\rangle(t)=p_0\sin t,
\qquad
\langle p\rangle(t)=p_0\cos t ,
\end{equation}
independently of $\kappa$.

Fluctuations are defined in the standard operator sense,
\begin{equation}
(\Delta A)^2
\equiv
\langle \hat A^2\rangle-\langle \hat A\rangle^2 ,
\end{equation}
with $\hat q=q$ and $\hat p=-i\kappa\partial_q$.
For the harmonic oscillator, the position and momentum variances evolve as
\begin{equation}
\begin{aligned}
(\Delta q)^2(t)
&=
(\Delta q)^2_0\cos^2 t
+
(\Delta p)^2_0\sin^2 t,
\\
(\Delta p)^2(t)
&=
(\Delta p)^2_0\cos^2 t
+
(\Delta q)^2_0\sin^2 t ,
\end{aligned}
\label{HO_variances}
\end{equation}
where $(\Delta q)_0$ and $(\Delta p)_0$ denote the initial variances.

The initial momentum variance is obtained directly from the operator
definition and can be written in amplitude--phase variables as
\begin{equation}
(\Delta p)^2_0
=
\kappa^2
\int dq\,\rho(q,0)\,
\bigl(\partial_q\ln R(q,0)\bigr)^2 ,
\label{Dp0_operator}
\end{equation}
while $(\Delta q)_0$ is fixed solely by the spatial profile of the
initial density $\rho(q,0)$.

For the polynomial-modified Gaussian initial data
\eqref{initialRq0}, the integrals can be evaluated explicitly, yielding
\begin{align}
(\Delta q)_0^2
=&
\frac{\sigma^2}{2}\,
\frac{
\varepsilon^4
+3\,\varepsilon^2\sigma^2
+\tfrac{15}{4}\sigma^4
}{
\varepsilon^4
+\varepsilon^2\sigma^2
+\tfrac34\sigma^4
},
\\
(\Delta p)_0^2
=&
\frac{\kappa^2}{2\sigma^2}\,
\frac{
\varepsilon^4
-\,\varepsilon^2\sigma^2
+\tfrac74\sigma^4
}{
\varepsilon^4
+\varepsilon^2\sigma^2
+\tfrac34\sigma^4
}.
\end{align}

Consequently, the uncertainty product satisfies
\begin{equation}
\Delta q(t)\,\Delta p(t)\ge\frac{|\kappa|}{2}
\end{equation}
at all times. For Gaussian initial profiles (corresponding to the limit
$\varepsilon/\sigma \to \infty$), the bound is saturated at specific
instants of time, such as $t=0$.
In the classical limit $|\kappa|\to0$, this lower bound vanishes, and the
point-particle limit may then be recovered by subsequently taking
$\sigma\to0$.

\subsection*{Interpretation}

This elementary example illustrates, in a fully controlled setting, the
conceptual viewpoint advocated throughout this work.

The starting point is entirely classical.
Given a Hamiltonian system, the dynamics of an ensemble is described by
the HJ equation for the principal function $S$ together
with the continuity equation for the density $\rho=R^2$.
At this level, the description is deterministic and requires no
probabilistic or quantum assumptions.

Introducing the complex wavefunction
$\psi = R\,e^{iS/\kappa}$
is a purely algebraic reparametrization of these classical variables.
For $|\kappa|\to0$, this representation remains well defined and simply
packages the classical ensemble data $(R,S)$ into a single complex
function.
In this limit, $R^2$ coincides with the classical density and $S$ reduces
to the HJ principal function.

A key observation is that, for finite $\kappa$, this reparametrization
renders the dynamics linear.
The coupled equations for $R$ and $S$ become exactly equivalent to a
single Schr\"odinger-type equation for $\psi$.
From this perspective, the Schr\"odinger equation does not represent an
independent postulate, but rather a linearized representation of the
underlying HJ ensemble dynamics.

Once the dynamics is cast in linear form, the familiar operator
structures of quantum mechanics emerge naturally.
Observables and their fluctuations admit a transparent definition in
terms of linear operators acting on $\psi$, while the underlying
ensemble dynamics remains fully encoded in $(R,S)$.
The harmonic oscillator provides a particularly clean benchmark, in
which the equivalence between the two descriptions can be demonstrated
explicitly.

\begin{figure*}[t]
\centering 
\includegraphics[width=\textwidth]{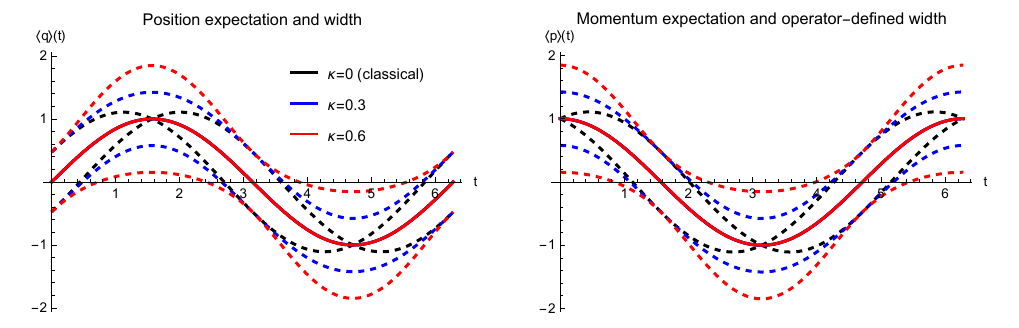}
\caption{
Visualization of the ensemble dynamics for the one-dimensional harmonic
oscillator.
\textbf{Left panel:} Time evolution of the position expectation value
$\langle q\rangle(t)$ (solid line) together with the width
$\langle q\rangle(t)\pm\Delta q(t)$ for several representative values of
$\kappa$.
The central trajectory is independent of $\kappa$ and coincides with the
classical solution, while the width increases with $\kappa$.
\textbf{Right panel:} Time evolution of the momentum expectation value
$\langle p\rangle(t)$ (solid line) together with the operator-defined
dispersion $\langle p\rangle(t)\pm\Delta p_{\mathrm{op}}(t)$.
Although $\langle p\rangle(t)$ is again purely classical, the width
captures the $\kappa$-dependent momentum fluctuations relevant for the
uncertainty relation.
The HJ momentum dispersion $\Delta p_{\mathrm{HJ}}(t)$,
which vanishes initially and follows the complementary $\sin^2 t$
behavior, is discussed in the Supplemental Material, section
\textit{HJ versus operator momentum fluctuations},  but not shown here. In the numerical illustrations we choose
$\varepsilon=\sigma=0.4$ and $p_0=1$,
which provide a smooth, localized initial ensemble and a clearly visible
classical trajectory.
\label{fig:HO_visual}
}
\end{figure*}

\subsection*{HJ versus operator momentum fluctuations}

Although the $(R,S)$ and $\psi$ descriptions are dynamically equivalent,
the linear HJS representation reveals a structural distinction between
two notions of momentum fluctuations that is usually not made explicit.
This distinction plays no role at the level of the equations of motion,
but becomes essential when discussing fluctuations, uncertainty
relations, and the operational meaning of observables.

Within the HJ description, one may define a classical
ensemble momentum dispersion,
\begin{equation}
(\Delta p)^2_{\mathrm{HJ}}
\equiv
\int dq\,\rho(q)\,
\bigl(\partial_q S(q)-\langle p\rangle\bigr)^2 ,
\end{equation}
which measures variations of the momentum flow field
$p(q)=\partial_q S(q)$ across configuration space.
For the initial data considered here, $S(q,0)=p_0 q$, this quantity
vanishes identically,
\begin{equation}
(\Delta p)_{\mathrm{HJ}}(0)=0,
\end{equation}
reflecting the fact that the HJ momentum field is spatially
constant.

By contrast, the momentum fluctuation entering the uncertainty relation
is defined through the linear momentum operator,
\begin{equation}
(\Delta p)^2_{\mathrm{op}}
\equiv
\langle \hat p^2\rangle-\langle \hat p\rangle^2,
\qquad
\hat p=-i\kappa\partial_q .
\end{equation}
For a general state $\psi=R e^{iS/\kappa}$, the two notions are related by
the exact identity
\begin{equation}
(\Delta p)^2_{\mathrm{op}}
=
(\Delta p)^2_{\mathrm{HJ}}
+
\kappa^2
\int dq\,\rho(q)\,
\bigl(\partial_q\ln R(q)\bigr)^2 .
\label{Dp_relation}
\end{equation}

For the specific initial data \eqref{initialRq0} considered here, the
evolution of the HJ momentum field yields
\[
(\Delta p)^2_{\mathrm{HJ}}(t)
=
(\Delta q)^2_0\,\sin^2 t .
\]
By contrast, the amplitude-gradient contribution is given explicitly by
\[
\kappa^2
\int dq\,\rho(q,t)\,
\bigl(\partial_q\ln R(q,t)\bigr)^2
=
(\Delta p)^2_0\,\cos^2 t .
\]

In the present example, although the HJ momentum field is
perfectly sharp initially, the operator-defined momentum dispersion is
nonzero  due to the spatial localization of the ensemble.
This illustrates explicitly that momentum fluctuations in the linear
HJS representation originate not only from variations of the classical
momentum flow, but also from the additional structure required to encode
the dynamics in a linear complex wavefunction.

Despite this distinction, the two descriptions are dynamically
equivalent.
The coupled evolution of $(R,S)$ and the linear Schr\"odinger evolution of
$\psi$ generate identical ensemble dynamics and identical expectation
values at all times.

\subsection*{Visualization}

To complement the analytical discussion in this section, we present a
compact visualization of the ensemble dynamics for the harmonic
oscillator.
The purpose of this figure is not to introduce new results, but to
provide an intuitive illustration of how the classical HJ 
dynamics, the linear HJS evolution, and the associated fluctuations are
organized within the same framework.

Figure~\ref{fig:HO_visual} shows the time evolution of the expectation
values $\langle q\rangle(t)$ and $\langle p\rangle(t)$ together with
their operator-defined fluctuations for several representative values of
the deformation parameter $\kappa$.
The expectation values follow the classical equations of motion exactly
and are independent of $\kappa$, while the widths encode the
$\kappa$-dependent spreading of the ensemble.

In the left panel, the position expectation value $\langle q\rangle(t)$
is shown together with the time-dependent width $\Delta q(t)$.
The central trajectory coincides with the classical solution, while the
envelope $\langle q\rangle(t)\pm\Delta q(t)$ illustrates how the ensemble
broadens as $\kappa$ is increased.
This spreading reflects the quantum potential term  $Q[R]$ in the modified
HJ equation and has no counterpart in the classical limit
$|\kappa|\to0$.

The right panel displays the momentum expectation value $\langle
p\rangle(t)$ together with the operator-defined momentum dispersion
$\Delta p_{\mathrm{op}}(t)$.
Although the expectation value again follows the classical trajectory
independently of $\kappa$, the width captures the momentum fluctuations
relevant for the uncertainty relation and for experimental measurements.
As discussed in the Supplemental Material, section
\textit{HJ versus operator momentum fluctuations}, this operator-defined fluctuation differs
conceptually from the dispersion of the HJ momentum field,
even though both descriptions are dynamically equivalent.

Taken together, the figure provides a direct visual confirmation of the
central message of this section:
the variables $(R,S)$ and the complex wavefunction $\psi$ encode the same
underlying dynamics, but the linear $\psi$--representation offers a
particularly transparent and operationally meaningful description of
fluctuations, uncertainties, and observable quantities.

\subsection*{Beyond the harmonic oscillator}

The absence of quantum corrections to the classical trajectory in the
present example is a special property of the harmonic oscillator.
For more general potentials, such as
\begin{equation}
V(q)=\frac12 q^2 + \lambda q^4 ,
\end{equation}
quantum effects generically modify both the effective trajectory and the
HJ momentum dispersion $(\Delta p)_{\mathrm{HJ}}$.

We do not pursue these more complicated situations here, as the purpose
of this section is solely to provide a simple, transparent, and fully
controlled benchmark example.

\section{Spin as an Internal Fiber in the HJS Framework}

The analysis in the main text focused on a single real degree of freedom.
We now clarify how finite-dimensional internal structures, such as spin,
enter the HJS framework and how they
couple to orbital motion.

\medskip
\noindent\textbf{Spin as an internal representation.}
Let $A$ label a finite-dimensional internal representation of a symmetry
group (e.g.\ $SU(2)$ for spin-$1/2$).
We write the complex wavefunction in polar form as
\begin{equation}
    \psi^{A}(q,t) \;=\; R(q,t)\,u^{A}(q,t)\,e^{iS(q,t)/\kappa},
\end{equation}
where $R\ge0$ is a scalar density, $S$ is a single action field shared by
all components, and $u^{A}$ is a normalized internal state,
$u^{\dagger}u=1$.
The internal structure is therefore carried entirely by $u^{A}$, while
the orbital dynamics is governed by $S$.

\medskip
\noindent\textbf{Coupling through an internal connection.}
Substituting this decomposition into the HJS equations shows that the
orbital velocity field depends on the effective momentum
\begin{equation}
    p_\mu^{\rm eff}
    \;=\;
    \partial_\mu S \;-\; \kappa\, a_\mu,
    \qquad
    a_\mu \equiv i\,u^\dagger \partial_\mu u .
\end{equation}
The internal degrees of freedom thus enter the orbital dynamics only
through the geometric connection $a_\mu$ on the internal fiber.
Importantly, dimensional consistency and Lorentz covariance require this
connection to appear multiplied by $\kappa$.

\medskip
\noindent\textbf{Decoupling in the $|\kappa|\to0$ limit.}
As a consequence, in the limit $|\kappa|\to0$ the internal connection
decouples from the HJ equation.
The internal state $u^{A}$ continues to evolve by parallel transport
along the characteristic curves, but it no longer feeds back into the
orbital dynamics.
This is a decoupling, not a collapse, of the internal degrees of freedom.

\medskip
\noindent\textbf{Deterministic dynamics at finite $\kappa$.}
For any finite value of $\kappa$, the coupled fields $(R,S,u)$ evolve under
first-order HJS evolution equations.
The resulting dynamics is continuous and deterministic, with the
internal fiber contributing geometric corrections to the effective
HJ flow.
The linear operator equation
\begin{equation}
    i\kappa\,\partial_t \psi^{A}
    \;=\; \hat{H}^{A}{}_{B}\,\psi^{B}
\end{equation}
is therefore a convenient linear representation of the same underlying
ensemble dynamics, rather than an independent postulate.

\medskip
In this way, spin appears in the HJS framework as an internal fiber degree
of freedom whose coupling to orbital motion is controlled by the
deformation parameter $\kappa$.
The theory remains deterministic for finite $\kappa$, while the classical
HJ limit corresponds to the geometric decoupling of the
internal structure.

This section is not a spin quantization scheme; rather, it shows that
the HJS embedding is compatible with the standard treatment of spin
and reproduces the usual linear dynamics once the internal fiber is
included.  The scalar analysis in the main text therefore extends to
spinful systems without conceptual modification.

\end{document}